\documentclass[journal,twocolumn]{IEEEtran}
\usepackage{color}
\usepackage{graphicx}
\usepackage{epstopdf}
\usepackage{amsmath}
\usepackage{amssymb}
\usepackage[english]{babel}
\usepackage{cite}
\usepackage{rotfloat}
\usepackage{mathtools}
\usepackage[justification=centering]{caption}
\usepackage{breqn}
\usepackage{amsmath}
\usepackage{makecell}
\usepackage{algorithm,algorithmic}
\usepackage{multirow}
\usepackage{subfigure}
\usepackage{booktabs}
\usepackage{colortbl}
\definecolor{kugray5}{RGB}{224,224,224}
\usepackage[export]{adjustbox}
\usepackage[normalem]{ulem}
\newcommand\rsout{\bgroup\markoverwith
	{\textcolor{red}{\rule[0.5ex]{2pt}{0.8pt}}}\ULon}



\makeatletter
\newcommand{\ALOOP}[1]{\ALC@it\algorithmicloop\ #1%
	\begin{ALC@loop}}
	\newcommand{\ENDALOOP}{\end{ALC@loop}\ALC@it\algorithmicendloop}

\makeatother

\usepackage{etoolbox}
\let\mybibitem\bibitem
\renewcommand{\bibitem}[1]{%
	\ifstrequal{#1}{nature}
	{\color{blue}\mybibitem{#1}}
	{\color{black}\mybibitem{#1}}%
}

\graphicspath{ {Figures/} }

\DeclareCaptionLabelSeparator{periodspace}{.\quad}

\renewcommand{\figurename}{Fig.}
\addto\captionsenglish{\renewcommand{\figurename}{Fig.}}
\newcommand\numberthis{\addtocounter{equation}{1}\tag{\theequation}}

\newcommand{\norm}[1]{\left\lVert#1\right\rVert} 
\newcommand{\eq}[1]{\begin{align*}#1\end{align*}} 
\newcommand{\eqn}[1]{\begin{align}#1\end{align}} 
\newcommand{\nt}[1]{\left(#1\right)} 
\newcommand{\nv}[1]{\left[#1\right]} 
\newcommand{\nn}[1]{\left\{#1\right\}} 
\newcommand{\abs}[1]{\left|#1\right|} 

\newcommand{\re}[1]{\mathfrak{R}{\left(#1\right)}}
\newcommand{\im}[1]{\mathfrak{I}{\left(#1\right)}}

\newcommand{\prob}{\mathbb{P}} 
\newcommand{\cdf}{\mathbf{\textit{F}}} 
 
\newcommand{\mean}[1]{\mathbb{E} \left\{#1\right\}}

\newcommand{\mQ}{\textbf{\textit{Q}}}
\newcommand{\mR}{\textbf{\textit{R}}}
\newcommand{\mH}{\textbf{\textit{H}}}

\newcommand{\setA}{\mathcal{A}} 
\newcommand{\setD}{\mathcal{D}} 
 
\newcommand{\setAtld}{\tilde{\setA}^{N_t}}
\newcommand{\setN}{\mathcal{N}(\vc)}

\newcommand{\vxb}{\textbf{\textit{x}}^{\star}}
\newcommand{\vc}{\textbf{\textit{c}}}

\newcommand{\vs}{\textbf{\textit{s}}}
\newcommand{\vx}{\textbf{\textit{x}}}
\newcommand{\vy}{\textbf{\textit{y}}}
\newcommand{\vr}{\textbf{\textit{r}}}

\newcommand{\vrd}{\textbf{\textit{r}}_d}

\newcommand{\vv}{\textbf{\textit{v}}}
\newcommand{\vu}{\textbf{\textit{u}}}
\newcommand{\vz}{\textbf{\textit{z}}} 
\newcommand{\vh}{\textbf{\textit{h}}}

\newcommand{\smt}{\sigma_t^2} 
\newcommand{\smv}{\sigma_v^2} 

\begin{document}

	\title{Groupwise Neighbor Examination for Tabu Search Detection in Large MIMO systems}
	\author{Nhan~Thanh Nguyen and~Kyungchun~Lee,~\IEEEmembership{Senior Member,~IEEE}
		\thanks{N. T. Nguyen and K. Lee are with the Department of Electrical and Information Engineering, Seoul National University of Science and Technology, Seoul 01811, Republic of Korea (e-mail: nhan.nguyen, kclee@seoultech.ac.kr).}
	}
	\maketitle
	\begin{abstract}
		In the conventional tabu search (TS) detection algorithm for multiple-input multiple-output (MIMO) systems, the metrics of all neighboring vectors are computed to determine the best one to move to. This strategy requires high computational complexity, especially in large MIMO systems with high-order modulation schemes such as 16- and 64-QAM signaling. This paper proposes a novel reduced-complexity TS detection algorithm called neighbor-grouped TS (NG-TS), which divides the neighbors into groups and finds the best neighbor by using a simplified cost function. Furthermore, based on the complexity analysis of NG-TS, we propose a channel ordering scheme that further reduces its complexity. Simulation results show that the proposed NG-TS with channel ordering can achieve up to 85$\%$ complexity reduction with respect to the conventional TS algorithm with no performance loss in both low- and higher-order modulation schemes.\looseness=-1
	\end{abstract}

	\begin{IEEEkeywords}
		Neighbor examination, Tabu search detection, massive MIMO, ordering schemes.
	\end{IEEEkeywords}
	\IEEEpeerreviewmaketitle
	
	\section{Introduction}
	
	Recently, the tabu search (TS) detector has been introduced as a complexity-efficient scheme for symbol detection in large multiple-input multiple-output (MIMO) systems. This is because it can perform very close to the maximum likelihood (ML) bound with far lower complexity compared to sphere decoding (SD) and fixed-complexity SD (FSD) \cite{srinidhi2011layered}, \cite{nguyen2019qr}. Several TS-based detection algorithms have been proposed for large MIMO systems, such as layered TS \cite{srinidhi2011layered}, reactive TS (RTS) \cite{srinidhi2009low}, random-restart reactive TS (R3TS) \cite{datta2010random}, and TS with early termination (ET) \cite{zhao2007tabu}. However, the performance-complexity trade-off has not been well optimized. Specifically, the performance of RTS is far from the optimum for higher-order modulation schemes such as 16- and 64-QAM \cite{srinidhi2009near}. In contrast, LTS and R3TS achieve improved bit-error-rate (BER) performance at the expense of increased complexity, especially for large MIMO and high-order QAMs. Furthermore, although TS with ET in \cite{zhao2007tabu} achieves complexity reduction by reducing the number of examined neighbors or by using a stopping criterion, it comes at the cost of performance loss. In \cite{feng2018low}, the RTS algorithm is further optimized by reducing the neighborhood search space, resulting in approximately $50 \%$ complexity reduction with almost no performance loss with respect to the conventional RTS algorithm. The QR-decomposition-aided TS (QR-TS) algorithm that uses an efficient metric-computation scheme to reduce the overall complexity of TS is introduced in \cite{nguyen2019qr}. The TS algorithm is also applied to various systems to improve performance. In \cite{jeong2019new}, the advantage of TS is used to find the initial solution with low complexity to mitigate the interference in generalized frequency-division multiplexing systems. In \cite{kayal2018dynamic}, Kayal et al. propose a TS-based dynamic thresholding approach to detect hard exudates in retinal images.
	
	In an $N_t \times N_r$ MIMO system with QPSK modulation, the number of neighbors in each searching iteration of TS algorithms can be up to $2N_t-1$ for QPSK, and $4N_t-1$ for 16- and 64-QAM \cite{nguyen2019qr}. Furthermore, a large MIMO system requires a large number of searching iterations to achieve near-ML performance. Therefore, the overall complexity of the TS-based detection algorithms becomes extremely high in large MIMO systems, and the most complexity arises from determining the best neighbors. This motivates the proposal of a novel TS detection algorithm, called neighbor-grouped TS (NG-TS) in this work. Our main contributions can be summarized as follows:
	\begin{itemize}
		\item By expanding the ML cost function, which is a function of a channel column vector, we show that among the neighbors corresponding to the same column norm of a channel matrix, the best one can be determined using a simplified cost function. This requires considerably less complexity than the scheme that employs the conventional cost function.
		
		\item By employing the simplified cost function, we develop the groupwise neighbor-examination scheme for the TS algorithms. Specifically, the neighbors are divided into groups, and the groups' best neighbors are compared to determine the final best neighbor. This scheme allows the best neighbor in each iteration to be found with much lower complexity than that of the sequential neighbor-examination approach used in prior TS schemes.

		\item Based on the complexity analysis of the NG-TS algorithm, we propose a channel ordering scheme for further complexity reduction. Our simulation results show that the proposed schemes can significantly reduce the complexity of the TS algorithm while fully preserving its BER performance. As a result, the performance--complexity tradeoff is improved.
	\end{itemize}

	\section{System Model}
	
	We consider the uplink of a multiuser MIMO system with $N_r$ receive antennas, where the total number of transmit antennas of all users is $N_t$. The received signal vector $\tilde{\vy}$ is given by\looseness=-1
	\eqn{
		\label{complex SM}
		\tilde{\vy} = \tilde{\mH} \tilde{\vs} + \tilde{\vv} ,
	}
	where $\tilde{\vs} = \nv{\tilde{s}_1, \tilde{s}_2, \ldots, \tilde{s}_{N_t}}^T$ is the vector of transmitted symbols. We assume that $\mean{\abs{s_i}^2} = \smt$, where $\smt$ is the average symbol power, and $\tilde{\vv}$ is a vector of independent and identically distributed (i.i.d.) additive white Gaussian noise (AWGN) samples $\tilde{v}_i \sim \mathcal{CN}(0,\smv)$. Furthermore, $\tilde{\mH}$ denotes an $N_r \times N_t$ channel matrix consisting of entries $h_{i,j}$, where $h_{i,j}$ represents the complex channel gain between the $j$th transmit antenna and the $i$th receive antenna, and $h_{i,j}$ is assumed to be an i.i.d. zero-mean complex Gaussian random variable. The transmitted symbols $\tilde{s}_i, i = 1, 2, \ldots, N_t,$ are independently drawn from a complex constellation $\tilde{\setA}$ of $Q$ points. The set of all possible transmitted vectors forms an $N_t$-dimensional complex constellation $\setAtld$ consisting of $Q^{N_t}$ vectors, i.e., $\tilde{\vs} \in \setAtld$.
	
	The complex signal model \eqref{complex SM} can be converted to an equivalent real signal model
	\eqn{
		\vy = \mH \vs + \vv, \label{real SM}
	}
	where $\vs, \vy, \vv,$ and $\mH$ given as
	\eq{
		\begin{bmatrix}
			\re{\tilde{\vs}}\\
			\im {\tilde{\vs}}
		\end{bmatrix}, 
		\begin{bmatrix}
			\re{\tilde{\vy}}\\
			\im {\tilde{\vy}}
		\end{bmatrix},
		\begin{bmatrix}
			\re{\tilde{\vv}}\\
			\im {\tilde{\vv}}
		\end{bmatrix}, \text{ and }
		\begin{bmatrix}
			\re {\tilde{\mH}}  &-\im {\tilde{\mH}}\\
			\im {\tilde{\mH}}  &\re {\tilde{\mH}}
		\end{bmatrix},
	}
	denote the $\nt{M \times 1}$-equivalent real transmitted signal vector, $\nt{N \times 1}$-equivalent real received signal, AWGN noise signal vectors, and the $(N \times M)$-equivalent real channel matrix, respectively, with $M = 2N_t, N = 2N_r$. Here, $\re {\cdot}$ and $\im {\cdot}$ denote the real and imaginary parts of a complex number or vector, respectively. In \eqref{real SM}, we have $\vs \in \setA^M$, where $\setA$ is the equivalent real-valued signal constellation set of $\tilde{\setA}$.
	
	For the description of the TS algorithm, we use the equivalent real-valued signal model in \eqref{real SM}. Subsequently, the maximum likelihood (ML) solution can be written as
	\eqn {
		\hat{\vs}_{ML} = \arg \min_{\vs \in \setA^{M}} \{ \phi(\vs) \}, \label{ML solution}
	}
	where $\phi(\vs) = \norm {\vy - \mH \vs}^2$ is the ML metric of $\vs$. The computational complexity of ML detection in \eqref{ML solution} is exponential with $M$ \cite{srinidhi2011layered}, which results in extremely high complexity for massive MIMO systems, where $M$ is very large.

	\section{Proposed NG-TS algorithm}
	
	\subsection{Conventional TS algorithm} 
	The TS algorithm starts with an initial candidate vector, which is often assumed to be the ZF solution $\vx_{ZF} = \ulcorner \mH^{\dagger} \vy \lrcorner$, where $\mH^{\dagger}$ is the pseudoinverse of $\mH$ and  $\ulcorner \cdot \lrcorner$ is element-wise quantization to the nearest point in $\mathcal{A}$. Then, it sequentially moves to $\mathcal{I}$ candidates for $\mathcal{I}$ iterations. In each iteration, the neighbors of the candidate $\vc$ are examined to find the best neighbor $\vxb$ with the smallest ML metric, i.e., $\vxb = \arg \min_{\vx \in \setN}  \{ \phi(\vx) \}$, where $\setN$ is the neighbor set of $\vc$. Here, the neighbors of $\vc$ are defined as the non-tabu vectors inside $\mathcal{A}^N$ with the smallest distance to $\vc$. 
		
	The ML metric $\phi (\vx)$ can be rewritten as $\phi (\vx) = \norm {\vu + \vh_d \delta_d}^2$ \cite{nguyen2019qr}, where $\vu = \vy - \mH \vc$, $\delta_d$ is the single nonzero element of $\vc - \vx = \nv {0, \ldots, 0, \delta_d, 0, \ldots, 0}^T$, and $\vh_d$ is the $d$th column of $\mH$. In this study, $d$ is called the \textit{difference position} of a neighbor, in which the candidate and its neighbor are different. For example, if the current candidate is $\vc=[1, -3, 1, 3]^T$, then $d=3$ is the difference position for $\vx = [1, -3, -1, 3]^T$ because $\vc $ and $\vx $ are only different at the third element. After the best neighbor is determined, it becomes the candidate in the next iteration, and then the best neighbor of a new candidate is determined. By using this iterative method, the best candidate visited for $\mathcal{I}$ iterations is chosen to be the final solution.\looseness=-1
	
	The overall complexity of the conventional TS algorithm, including the complexity involved in the initialization and iterative searching process,  can be given as \cite{nguyen2019qr}
	\begin{align*}
		\mathcal{C}_{\text {Conv. TS}} = \underbrace{\frac{2N^3}{3} + 3N^2 + \frac{N}{3}}_{\text {initialization}} + \underbrace{\mathcal{I} (4\bar{L} N - 2)}_{\text {iterative search}} \numberthis \label{comp_conv_TS},
	\end{align*}
	where $\bar{L}$ is the average number of neighbors in an iteration. It is worth noting that most of the complexity of TS algorithms arises from the process of finding the best neighbors during searching iterations, especially in large MIMO systems where $\bar{L}$ is large and a very large $\mathcal{I}$ is required to achieve near-optimal performance. Motivated by this, in the next subsection, we propose a novel scheme to reduce the complexity of the neighbor examination in TS algorithms. 
	
	\subsection{NG-TS algorithm}
	
For the efficient computation of neighbors' metrics, the work of \cite{nguyen2019qr} introduces a reduced cost function
	\eq{
		\phi (\vx) = \norm{\vz + \vrd \delta_d}^2, \numberthis \label{metric_QR_0}
	}
	where $\vz = \mQ^T \vy-\mR \vc$, the unitary matrix $\mQ$, and upper triangular matrix $\mR$ are obtained by the QR decomposition of $\mH$, i.e., $\mH=\mQ \mR$, and $\vrd$ is the $d$th column of $\mR$. In this work, we employ \eqref{metric_QR_0} to derive the proposed NG-TS algorithm.
	\subsubsection{Neighbor grouping}
	In the proposed NG-TS algorithm, the neighbor set $\mathcal{N} (\vc)$ is divided into groups, and the best neighbor of each group is determined based on a simplified cost metric function. Let $\vx_l$ and $d_l$ be the $l$th neighbor in $\setN$ and its difference position, $l = 1, 2, \ldots, L$, where $L$ is the number of neighboring vectors in $\mathcal{N} (\vc)$. We note that the distances between the candidate $\vc$ to all neighbors are the same, i.e., $\abs{\delta} = \abs{\delta_{d_1}} = \ldots = \abs{\delta_{d_L}}$. By expanding $\phi(\vx_l)$ in \eqref{metric_QR_0}, we obtain\looseness=-1
	\begin{align*}
	\phi(\vx_l) = \norm {\vz}^2 + \abs{\delta}^2 \norm {\vr_{d_l}}^2 + 2 \delta_{d_l} \vz^T \vr_{d_l}. \numberthis \label{metric_QR_1}
	\end{align*}
	It is observed from \eqref{metric_QR_1} that among the neighbors having the same value for $\norm {\vr_{d_l}}$, the one with smallest $\delta_{d_l} \vz^T \vr_{d_l}$ has the smallest ML metric. Let $\mathcal{G}_k$ be a group of neighbors having the same value $\norm {\vr_{d_l}}$, i.e.,\looseness=-1
	\begin{align*}
	\mathcal{G}_k = \left\{ \vx_i \in \setN: \norm {\vr_{d_i}} = \eta_k \right\}, \numberthis \label{group}
	\end{align*}
	where $\eta_k$ is one of the column norms of $\mR$, i.e., $\eta_k \in \left\{ \norm {\vr_1}, \norm {\vr_2}, \ldots, \norm {\vr_N} \right\}$. Hence, the best neighbor $\vx_{\mathcal{G}_k}^{\star}$ in group $\mathcal{G}_k$ can be found with a simplified cost function as follows:\looseness=-1
	\begin{align*}
	\vx_{\mathcal{G}_k}^{\star} 
	= \arg \min_{\vx \in \mathcal{G}_k} \left\{\text{sign}(\delta_{d_l}) \gamma_{d_l}\right\}, \numberthis \label{bestnb_group_2}
	\end{align*}
	where $\gamma_{d_l} = \vz^T \vr_{d_l}$.
	
	The number of neighboring vectors in each group can be obtained from \eqref{group}, with the note that in the real signal model, $\norm{\vr_n} = \norm {\vr_{n+N_t}}, n=1,\ldots,N_t$, and $\vx_n$ and $\vx_j$ are in the same group when $d_n = d_j$. In QPSK, because each symbol in the alphabet $\left\{-1, 1\right\}$ only has one neighboring symbol, each group has a maximum of two vectors $\vx_{n}$ and $\vx_{n + N_t}$. By contrast, in higher-order modulation schemes such as 16-QAM and 64-QAM, whose alphabets are $\left\{-3, -1, 1, 3\right\}$ and $\left\{-7, -5, -3, -1, 1, 3, 5, 7\right\}$, respectively, each symbol has at most two neighboring symbols. Therefore, there is a maximum of four neighbors in each group, including two pairs of neighbors with the same difference positions. For example, with $M=4$ and 64-QAM modulation, a group can be formed by four vectors $[-7,-7,5,5]^T, [-7,-3,5,5]^T, [-7,-5,5,7]^T, [-7,-5,5,3]^T$, which are a subset of the neighbor set of $\vc = \left[-7,-5,5,5\right]^T$ with the difference positions $\left\{2, 4\right\}$.

	\subsubsection{Complexity of finding the groups' best neighbors}
	
	Determining the best neighbor for each group requires the computation of $\gamma_{d_l} = \vz^T \vr_{d_l} = \sum_{n=1}^{N} z_n r_{n,d_l}$. However, its complexity is less than that of a multiplication between two $N-$element vectors. This is because over two successive searching iterations, only a subset of elements of $\vz$ is updated as follows:
	\begin{align*}
	\vz_{\nn{i+1}} 
	&= \mQ^T \vy -\mR \vc_{\nn{i+1}} = \mQ^T \vy - \mR \left( \vc_{\nn{i}} - \Delta \vx_{\nn{i}} \right) \\
	&= \vz_{\nn{i}} + \mR \Delta \vx_{\nn{i}}  = \vz_{\nn{i}} + \vr_{d^{\star}} \delta_{d^{\star}}\\
	&= \vz_{\nn{i}} + \nv{r_{1,d^{\star}} \delta_{d^{\star}}, \ldots, r_{d^{\star},d^{\star}} \delta_{d^{\star}}, 0, \ldots, 0}^T,  \numberthis \label{sub_u_1}
	\end{align*}
	where the subscript $\left\{i\right\}$ represents the $i$th iteration\footnote{For notational convenience, we omit the interation index $i$ if it does not cause any confusion.}, and $\Delta \vx_{\nn{i}} = \vc_{\nn{{i}}} - \vc_{\nn{{i+1}}} = \left[0, \ldots, 0, \delta_{d^{\star}}, 0, \ldots, 0 \right]^T$ only has one non-zero element at the $d^{\star}$th position because $\vc_{\nn{{i}}} $ and $\vc_{\nn{{i+1}}}$ are neighbors of each other. 
	It is observed from \eqref{sub_u_1} that over two successive iterations, only the first $d^{\star}$ elements of $\vz$ need to be updated. Then,  $\gamma_{d_l}$ can be computed as
	\begin{align*}
	\gamma_{d_l} = \vz^T \vr_{d_l} =
	\begin{cases*}
	\tilde{\gamma}_{d_l} + \sum_{n=1}^{d^{\star}} z_n r_{n,{d_l}} , {d_l} \geq d^{\star}\\
	\sum_{n=1}^{{d_l}} z_n r_{n,{d_l}} , {d_l} < d^{\star}
	\end{cases*}, \numberthis \label{gamma_1}
	\end{align*}
	where $\tilde{\gamma}_{d_l} = \sum_{n=d^{\star}+1}^{d_l} z_n r_{n,{d_l}}$ was already computed in the previous iteration. Therefore, the computational complexity required in \eqref{gamma_1} to examine a neighbor is only $\min \left\{d_l, d^{\star} \right\}$ multiplications and $\min \left\{d_l-1, d^{\star} \right\}$ additions.
\begin{algorithm}[t]
	\caption{NG-TS Detection}
	\label{algorithm:NG-TS}
	\begin{algorithmic}[1]
		\REQUIRE $\mH  = \nv{\vh_1,\vh_2,\ldots,\vh_N}$, $\vy$
		\ENSURE $\hat{\vs}_{TS}$.
		
		\STATE {Obtain $\mQ$ and $\mR$ by QR-decomposition of $\mH$.}
		\STATE {Compute $f_m = \abs{\delta}^2 \norm{\vr_{m}}^2, m=1,2,\ldots,M $.}
		\STATE {Order the columns of $\mH$ in increasing order of $f_m,m=1,2,\ldots,M$, if channel ordering is applied.}
		
		\STATE {$\vx_{ZF} = \ulcorner \mR^{-1} \mQ^T \vy \lrcorner$}
		\STATE {$\vc = \vx_{ZF}, \vz = \mQ^T \vy - \mR \vc$} 
		\STATE {$\hat{\vs}_{TS} = \vc, \phi \nt{\hat{\vs}_{TS}} = \phi \nt{\vc} = \norm {\vz}^2$}
		\STATE {Push $\vc$ to the tabu list.} 
		
		\FOR {$i = 1$ to $\mathcal{I}$}
		\STATE {Find the neighbor set $\setN = \left\{ \vx_1, \ldots, \vx_L \right\}$ and the set of difference positions $\setD = \left\{ d_1, \ldots, d_L \right\}$.}
		
		\STATE{Divide $\setN$ into groups $\mathcal{G}_1, \ldots, \mathcal{G}_K$ based on \eqref{group}}.
		\FOR{$k = 1 \rightarrow K$}
		\FOR{$l = 1 \rightarrow L_k$}
		\STATE{Set $\vx$ and $d$ to the $l$th neighbor in $\mathcal{G}_k$ and its difference position}.
		\STATE{$\alpha_l = \text{sign} (\delta_d) \gamma_d$, where $\gamma_d = \vz^T \vr_{d}$}
		\ENDFOR
		
		\STATE {$\hat{l} = \arg \min_{l} \left\{\alpha_l\right\}$}
		\STATE {Set $\vx_{\mathcal{G}_k}^{\star}$ to the $\hat{l}$th neighbor in $\mathcal{G}_k$.}
		\STATE {Set $d_k^{\star}$ to the difference position of $\vx_{\mathcal{G}_k}^{\star}$.}
		\STATE {$\beta_k = 2 \delta_{d_k^{\star}} \gamma_{d_k^{\star}} + f_{d_k^{\star}}$}
		\ENDFOR
		
		\STATE {$\hat{k} = \arg \min_{k}  \left\{\beta_k\right\}$}
		\STATE {$\vxb = \vx_{\mathcal{G}_{\hat{k}}}^{\star}$}
		
		\STATE {Move to the best neighbor, $\vc = \vxb$, and update $\vz$.}
		\STATE {Update $\hat{\vs}_{TS} = \vc,\phi \nt{\hat{\vs}_{TS}} =  \phi \nt{\vc}$ if $\phi \nt{\vc} < \phi \nt{\hat{\vs}_{TS}}$.}
		
		\STATE {Release the first element in the tabu list if it is full.}
		\STATE {Push $\vc$ to the tabu list and update its length.}
		\ENDFOR
		\STATE {The final solution is the best solution $\hat{\vs}_{TS}$ found so far.}
	\end{algorithmic}
\end{algorithm}
	\subsubsection{Channel ordering}
	The complexity to find the best neighbor of each group can be further reduced by statistically minimizing $d^{\star}$ in searching iterations because the complexity of \eqref{gamma_1} increases with $\min \left\{d_l, d^{\star} \right\}$ and $\min \left\{d_l-1, d^{\star} \right\}$. We note that the average metric of $\vx^{\star}$ can be expressed as \cite{nguyen2019qr}
	\begin{align*}
	\mean{\phi (\vx^{\star})} = N \sigma_v^2 + \delta^2 \norm {\vh_{d^{\star}}}^2,
	\end{align*}
	where $d^{\star}$ is not only the column index but also the difference position of $\vx^{\star}$. Therefore, if we order the channel matrix such that $\norm{\vh_1}^2 \leq \ldots \leq \norm{\vh_N}^2$, there is a larger chance that the best neighbor has a small difference position. As a result, $d^{\star}$ can decrease. This motivates the ordering of channel columns in the increasing order of their norms to reduce the computational complexity of computing $\gamma_d$ in \eqref{gamma_1}.
	\begin{figure}[t]
		\centering
		\includegraphics[scale=0.55]{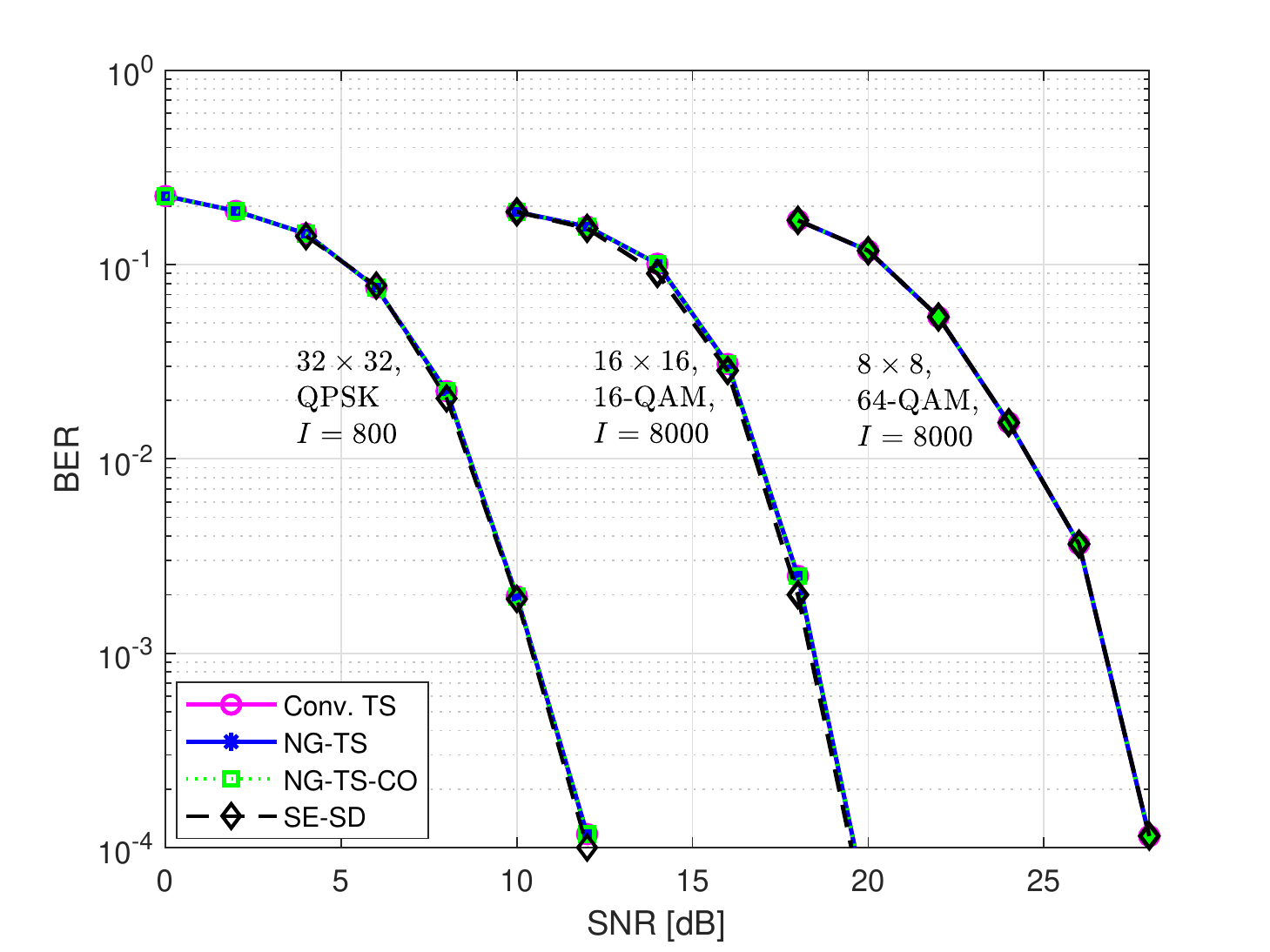}
		\caption{BER performance of the proposed schemes in comparison with those of the conventional TS and SE-SD.}
		\label{fig:ber_8x8}
	\end{figure}
	\begin{figure*}[t]
		\centering
		\includegraphics[scale=0.65]{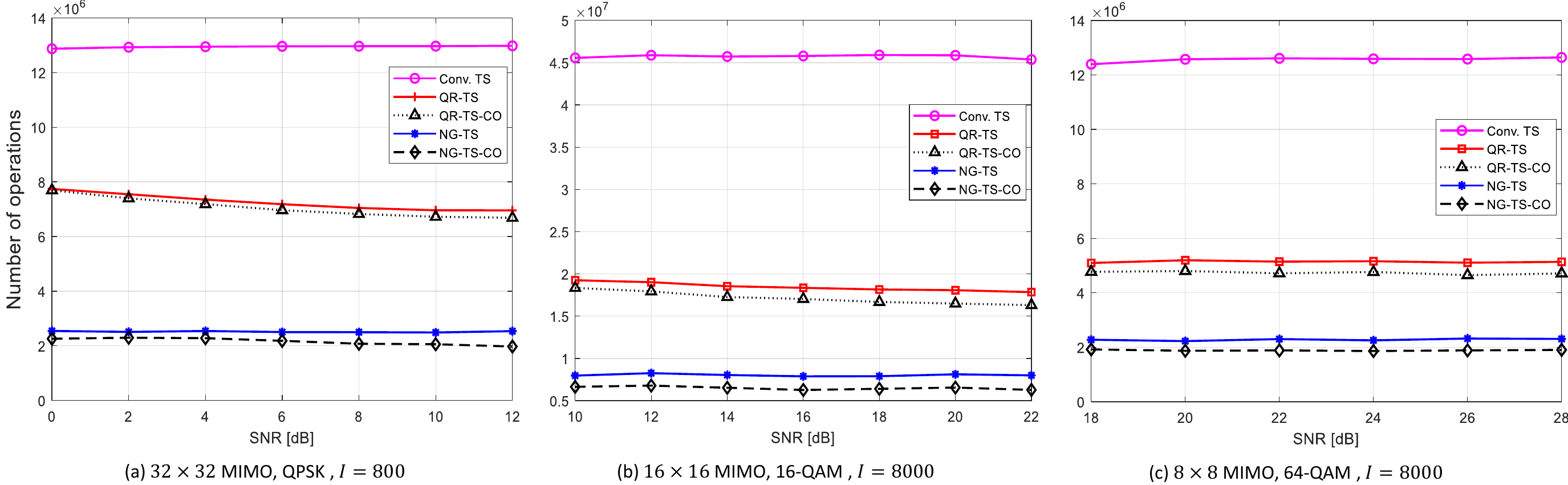}
		\caption{Average computational complexities of the proposed schemes, namely NG-TS and NG-TS-CO, in comparison with those of the conventional TS and QR-TS algorithms. The computational complexity is computed as total number of required multiplications and additions.}
		\label{fig:comp_fig}
	\end{figure*}
	
	\subsubsection{Finding the final best neighbor}
	Let $K$ be the number of groups of neighbors, and let $\vx_{\mathcal{G}_k}^{\star}$ and $d_k^{\star}$ be the best neighbors in group $\mathcal{G}_k$ and its difference position, respectively. Once $\vx_{\mathcal{G}_1}^{\star}, \ldots, \vx_{\mathcal{G}_K}^{\star}$ are found, the final best neighbor is set to $\vxb = \vx_{\mathcal{G}_{\hat{k}}}^{\star}$ such that\looseness=-1
	\begin{align*}
	\hat{k} 
	&= \arg \min_{k}  \phi(\vx_{\mathcal{G}_k}^{\star})\\
	&= \arg \min_{k}  \left\{\norm {\vz}^2 + 2 \delta_{d_k^{\star}} \gamma_{d_k^{\star}} + \abs{\delta}^2 \norm {\vr_{{d_k^{\star}}}}^2\right\}\\
	&= \arg \min_{k}  \left\{2 \delta_{d_k^{\star}} \gamma_{d_k^{\star}} + \abs{\delta}^2 \norm {\vr_{d_k^{\star}}}^2\right\},  \numberthis \label{final_bestnb}
	\end{align*}
	where the last equation is obtained by the fact that $\norm {\vz}^2$ is the same for all neighbors in each searching iteration. We note that $\gamma_{d_k^{\star}}$ in \eqref{final_bestnb} was already computed to search for the groups' best neighbors. Furthermore, $\abs{\delta}^2 \norm {\vr_{d_k^{\star}}}^2$ in \eqref{final_bestnb} only depends on the constant $\abs{\delta}^2$ and a column of $\mR$, which remain unchanged over $\mathcal{I}$ searching iterations. Therefore, $\abs{\delta}^2 \norm {\vr_{d_k^{\star}}}^2$ needs to be computed only once outside the searching iterations. As a result, the computational complexity to determine the final best neighbor is relatively low.\looseness=-1

	The proposed NG-TS algorithm is summarized in Algorithm \ref{algorithm:NG-TS}. In step 1, matrices $\mQ$ and $\mR$ are obtained by the QR decomposition of $\mH$. In step 2, $f_m = \abs{\delta}^2 \norm{\vr_{m}}^2, m=1,2,\ldots,M, $ are computed to be used in steps 3, 10, and 19, noting that $\norm{\vr_{m}}^2=\norm{\vh_{m}}^2, m=1,2,\ldots,M $. Step 4 computes the initial solution $\vx_{ZF} = \ulcorner \mH^{\dagger} \vy \lrcorner$. Then, steps 5--7 assign $\vx_{ZF}$ to the current candidate $\vc$, compute $\vz$, and initialize the solution $\hat{\vs}_{TS}$, which is then pushed to the tabu list. In step 10, $\setN$ is divided into $K$ groups of $L_k$, $k=1,\ldots,K$, neighboring vectors, which allows finding $K$ groups' best neighbors with a simplified cost function in steps 16 and 17. Then, the best neighbors of the $K$ groups are compared to determine the final best neighbor in steps 20 and 21. The following steps are for updating the best solution and the tabu list, and then conclude the final solution after $\mathcal{I}$ searching iterations.\looseness=-1
	
	\subsubsection{Computational complexity of the NG-TS detection algorithm}
	\begin{table}[t]
	\renewcommand{\arraystretch}{1.3}
	\caption{Computational complexity of Algorithm 1}
	\label{tab_comp_DSN}
	\centering
	\begin{tabular}{c|c|c}
		\hline
		Step  & Number of multiplications & Number of additions \\
		\hline
		\hline
		
		1  & $2N^3/3$ & $2N^3/3 $ \\
		\hline
		
		2 & $N^2/4  + 1$ & $N^2/4  - N/2$ \\
		\hline
		
		4 & $N^2/2 + N/2$ & $N^2/2 - N/2$ \\
		\hline
		
		5 & $3N^2/2 + N/2$ & $3N^2/2 - N/2$ \\
		\hline
		
		6 & $N$ & $N - 1$ \\
		\hline
		
		12--20 & $\makecell{2K +  \sum_{k=1}^{K} \sum_{l=1}^{L_k} (1 + \eta) }$ & $\makecell{K + \sum_{k=1}^{K}\sum_{l=1}^{L_k} \eta }$ \\
		\hline

	\end{tabular}
	\end{table}
	
	The computational complexity of the proposed NG-TS detection algorithm is presented in Table I with the assumption $M=N$. We assume that the QR Householder method is used to perform QR decomposition in step 1 of Algorithm 1. To solve $\vx_{ZF}$ in step 4, the backward substitution method is used. In step 14, the computation of $\gamma_d$ requires $\min \left\{d_l, d^{\star} \right\}$ multiplications and $\min \left\{d_l-1, d^{\star} \right\}$ additions, as discussed in Section III-B-2. For simplicity, we assume $\min \left\{d_l - 1, d^{\star} \right\} \approx \min \left\{d_l, d^{\star} \right\} = \epsilon$.\footnote{I've changed $\eta$ to $\epsilon$ to avoid confusing with $\eta_k$ in (7).} Therefore, computing $\alpha_l$ requires $1 + \epsilon$ multiplications and $\epsilon$ additions. Furthermore, computing $\beta_k$ in step 19 requires only 2 multiplications and 1 addition because $f_{d_{k^{\star}}}$ is already computed in step 2. Therefore, the total complexity of steps 12--20 is $2K + \sum_{k=1}^{K} \sum_{l=1}^{L_k} \left( 1 + \epsilon \right)$ multiplications and $K + \sum_{k=1}^{K} \sum_{l=1}^{L_k} \epsilon $ additions. Then, the average complexity required in an iteration of the NG-TS algorithm can be expressed as $\mathcal{C}_{iter} \approx 3\bar{K} + \bar{L} + 2 \varepsilon$, where $\bar{K}$ and $\bar{L}$ are the average values of $K$ and $L$ over all iterations, respectively, and 
	\begin{align*}
		\varepsilon &= \mean {\sum_{k=1}^{K} \sum_{l=1}^{L_k} \epsilon} = \sum_{k=1}^{K} \sum_{l=1}^{L_k} \mean {\epsilon}\\ 
		& = \bar{L} \mean {\epsilon} = \bar{L} \sum_{i=1}^{N}  i \left( \cdf_{\epsilon} [i] - \cdf_{\epsilon} [i - 1] \right). \numberthis \label{mean_sum}
	\end{align*}
	Here, $\mean{\cdot}$ represents an expected value, and $\cdf_{X} [\cdot]$ denotes the cumulative distribution function (CDF) of a random variable $X$. Recall that $\epsilon = \min \left\{d_l, d^{\star} \right\}$ with both $d_l$ and $d^{\star}$ being independent discrete random variables uniformly distributed on $[1,N]$. We have $\cdf_{d_l} [i] = \cdf_{d^{\star}} [i] = \frac{i}{N}$. Hence, $\cdf_{\epsilon} [i]$ can be expressed as
	\begin{align*}
		\cdf_{\epsilon} [i] &= \prob \left\{ \min \left\{d_l, d^{\star} \right\} \leq i \right\} = 1 - (1 - \cdf_{d_l} [i])(1 - \cdf_{d^{\star}} [i])\\
		&= 1 - \left( 1 - \frac{i}{N} \right)^2 =  \frac{i(2N-i)}{N^2},  \numberthis \label{cdf_eta} 
	\end{align*}
	where $\prob \left\{ \cdot \right\}$ denotes a probability. From \eqref{mean_sum} and \eqref{cdf_eta}, we have $\varepsilon = \bar{L} \sum_{i=1}^{N} \frac{(2N+1) i - 2i^2}{N^2} = \bar{L} \left(\frac{N}{3} + \frac{1}{2} + \frac{1}{6N}\right)$, and $\mathcal{C}_{\text {iter}}$ can be expressed as
	\begin{align*}
		\mathcal{C}_{\text {iter}} \approx 3\bar{K} + 2 \bar{L} + \frac{2 \bar{L}N}{3} + \frac{\bar{L}}{3N}. \numberthis \label{comp_iter_2}
	\end{align*}
	From Table I and \eqref{comp_iter_2}, the overall complexity of the NG-TS algorithm is given as
	\begin{align*}
		\mathcal{C}_{\text{NG-TS}} &\approx \underbrace{\frac{4N^3}{3} + \frac{9N^2}{2} + \frac{3N}{2}}_{\text {initialization}} + \underbrace{\mathcal{I} \left( 3\bar{K} + 2 \bar{L} + \frac{2 \bar{L} N}{3} + \frac{\bar{L}}{3N} \right)}_{\text {iterative search}}. \numberthis \label{comp_NGTS}
	\end{align*}
	
	By comparing \eqref{comp_NGTS} to \eqref{comp_conv_TS}, it is observed that the complexity required for the initialization of the NG-TS algorithm is greater than that of the conventional TS algorithm. However, in large MIMO systems, $\mathcal{I} \gg N$ is required to achieve near-optimal performance \cite{nguyen2019qr}. Therefore, the complexity required in the iterative searching process dominates the overall complexity. Furthermore, we note that $\bar{L} \leq N-1$ and $\bar{K} \leq 2$ for BPSK/QPSK, and $\bar{L} \leq 2N-1$ and $\bar{K} \leq 4$ for higher-order modulation schemes, such as 16- and 64-QAM, as discussed in Section I for $\bar{L}$ and in Section II-B-1 for $\bar{K}$. Therefore, from \eqref{comp_conv_TS} and \eqref{comp_NGTS}, it is clear that the complexity of the proposed NG-TS algorithm is significantly lower than that of the conventional TS algorithm. This will be numerically verified in the next section.

	\section{Simulation Results}
	
	In our simulations, each channel coefficient is assumed to be an i.i.d. zero-mean complex Gaussian random variable with a variance of $1/2$ per dimension, and the signal-to-noise ratio (SNR) is set to $ \smt / \smv $. In both Figs. 1 and 2, we consider $N_t=N_r=\left\{32, 16, 8\right\}$ with QPSK, 16-QAM, and 64-QAM, respectively. For those systems, $\mathcal{I}$ is set to $800, 8000$, and $8000$, respectively, which guarantees that the conventional TS and NG-TS algorithms perform approximately the same as the Schnorr-Euchner SD (SE-SD) decoder \cite{agrell2002closest}. Furthermore, the length of the tabu list is set to $P = \mathcal{I}/2$ so that TS algorithms achieve approximately the optimal performance \cite{zhao2007tabu, nguyen2019qr}.
	
	In Fig. \ref{fig:ber_8x8}, we compare the BER performance of the proposed NG-TS algorithms to those of the conventional TS and SE-SD algorithm. It is shown that in all the three considered systems, the proposed NG-TS and NG-TS with channel ordering (NG-TS-CO) schemes totally preserve the BER performance of the conventional TS algorithm. Furthermore, with the chosen values of $\mathcal{I}$ and $P$, the performances of TS algorithms are close to those of SE-SD.\looseness=-1
	
	In Fig. \ref{fig:comp_fig}, the computational complexities of the NG-TS, conventional TS, QR-TS, and QR-TS with channel ordering (QR-TS-CO) algorithms are compared. The same values of $\mathcal{I}$ and $P$ as in Fig. \ref{fig:ber_8x8} are used. It is shown that in all the considered environments, the complexity reduction ratios of NG-TS and NG-TS-CO compared with the conventional TS algorithm are approximately $80 \%$ and $85 \%$, respectively. Furthermore, the complexities of NG-TS are approximately $55\% -  70 \%$ lower than those of QR-TS.
	
	It has been shown in \cite{nguyen2019qr} that the QR-TS totally preserves the performance of the conventional TS and achieves the performances of SE-SD \cite{agrell2002closest}, layered TS \cite{srinidhi2011layered}, and $K-$best SD \cite{guo2006algorithm} algorithms with considerably lower complexities. Therefore, based on the comparisons of the performances and complexities of the conventional TS, QR-TS, and NG-TS in Figs. 1 and 2, it is clear that the proposed NG-TS achieves the improved performance-complexity trade-off compared to the conventional TS, QR-TS, SE-SD, LTS, and KSD.

	\section{Conclusion}
	In this paper, we propose a novel NG-TS algorithm as a solution for the neighbor examination of TS-based symbol detection. The proposed algorithm allows finding the best neighbor with low complexity by using a simplified cost function in a groupwise manner. In addition, a channel ordering scheme is proposed to further optimize the complexity of the proposed NG-TS algorithm. Simulation results show that the proposed NG-TS schemes can achieve up to 85$\%$ complexity reduction with respect to the conventional TS algorithm without any performance loss. We note that the proposed algorithm can be applied to other existing TS-based detection algorithms \cite{srinidhi2011layered, srinidhi2009low, datta2010random, zhao2007tabu, nguyen2019qr, srinidhi2009near} as an efficient neighbor examination scheme to reduce their computational complexities.\looseness=-1
	
	\bibliographystyle{IEEEtran}
	\bibliography{IEEEabrv,Bibliography}
	
\end{document}